%% file: main.tex
\documentclass[acmsmall]{acmart}
\AtBeginDocument{%
  }

\setcopyright{cc}
\setcctype{by-nc-nd}
\acmJournal{PACMHCI}
\acmYear{2027} \acmVolume{11} \acmNumber{1} \acmArticle{GROUP45}
\acmMonth{1} \acmDOI{10.1145/3859431}

\begin{document}

\title{Supporting Community Members Navigating Reentry with Revisable Future Envisioning}

%

\author{Modhumonty Das}
\email{mod53@pitt.edu}
\affiliation{ 
\department{School of Computing and Information}
  \institution{University of Pittsburgh}
  \city{Pittsburgh}
  \state{PA}
  \country{USA}
}

\author{Nathan Morelli}
\email{nbm27@pitt.edu}
\affiliation{ 
\department{School of Computing and Information}
  \institution{University of Pittsburgh}
  \city{Pittsburgh}
  \state{PA}
  \country{USA}
}

\author{Aakash Gautam}
\email{aakash@pitt.edu}
\affiliation{ 
\department{School of Computing and Information}
  \institution{University of Pittsburgh}
  \city{Pittsburgh}
  \state{PA}
  \country{USA}
}


\begin{abstract}
Future-planning tools often treat agency as individual and leave out the institutional conditions that shape how aspirations become action. We examine this tension in the context of reentry from incarceration and recovery from substance use disorder, through a partnership with a nonprofit organization. 
Building on participants’ physical vision-board practices, we designed a digital future-envisioning tool that helps users represent current circumstances, desired futures, and pathways that connect goals to concrete resources. 
We draw on data from three workshops involving 18 participants and follow-up interviews with 10 participants to examine how they used future envisioning within a fragmented support system. Participants described reentry and recovery as ongoing processes in which the everyday work of staying stable was central to progress. Envisioning futures supported this work by surfacing constraints and interdependencies, by reframing personal strengths and institutional supports as assets, and by helping participants move from abstract hopes toward articulated goals and contingent plans. We theorize revision as maintenance work and conclude with design implications for community-embedded interactive systems that support revisable, dependency-aware planning.
\end{abstract}

\ccsdesc[500]{Human-centered computing~Empirical studies in HCI}
\ccsdesc[300]{Human-centered computing~Empirical studies in collaborative and social computing}

\keywords{future envisioning, reentry, goal-setting, vision board, planning, maintenance, repair, infrastructure}


\received{31 March 2026}
\received[revised]{17 July 2026}
\received[accepted]{17 September 2026}

\maketitle
\input{introduction}
\input{litReviewUpdated}

\input{methodology}
\input{findings}
\input{discussion}
\input{conclusion}

\begin{acks}
We are grateful to the participants for sharing their experiences with us and for their willingness to try a tool with no guarantee that it would support their journeys. We thank West End P.O.W.E.R. for their collaboration and support throughout this work, and Rachel Siddiqui for championing our ongoing partnership. 
We also thank Unsplash for free API access and the anonymous reviewers for their constructive feedback. This work was supported in part by the University of Pittsburgh's Pitt Momentum Funds (2023--2024). 

We acknowledge the use of generative AI tools in the preparation of this work. ChatGPT (GPT 5.2) and Claude (Sonnet 4.5, Opus 5.5) were used to edit, summarize, and refine text.
\end{acks}

\bibliographystyle{ACM-Reference-Format}
\bibliography{Reentry_Support}



\end{document}

%% file: introduction.tex
\section{Introduction}
\label{introduction}


Reentry, the process of rejoining society after incarceration, recovery from substance use disorder, or both, is often marked by instability.
Community members navigating reentry have to coordinate across fragmented institutional systems as they return to housing, employment, relationships, and civic life \cite{Petersilia_2001, gautam_enhancing_2024, riggs2015reflections, phelan2022work}.

Each of these systems comes with deadlines, documentation requirements, shifting eligibility rules, and a risk that delayed responses will cascade into larger setbacks \cite{halushka2019runaround, visher2017evaluating, notley2013experience}.
Even ``basic'' goals like {employment} or housing are challenging because they rely on fragile conditions like reliable transportation, stable routines, and relationships with case managers, family, and peers.
Progress is rarely linear \cite{halushka2019runaround, dong2018competing, ortiz2019system}.
Plans are negotiated again and again as circumstances change and as institutional responses open and close pathways.

This reality does not align with how most planning and goal-setting tools are designed. 
Digital tools often treat goals as individual targets pursued through linear steps, and personal informatics systems in HCI have often assumed stable baselines and consistent progress \cite{li2010stage}.
More recent scholarship shows that people's goals shift between the abstract and the concrete over time \cite{niess2018supporting}, and a longitudinal study of goal setting found that complex goals require iterative selection, simplification, and adjustment \cite{agapie2022longitudinal}.
We build on these insights and argue that goal-setting systems must account for how goals evolve, how they depend on one another, and the ongoing work of keeping plans viable.

For community members navigating reentry, this work involves aligning everyday routines with external requirements, such as probation and parole restrictions, program attendance, and administrative timelines.
This coordination has been studied as articulation work in sociotechnical systems (e.g., \cite{strauss1988articulation, schmidt1992taking}), as the gap between plans and situated actions \cite{suchman1987plans}, and as repair and care work managing breakdowns \cite{jackson2014repair, karusala2023care, de2017matters}.
From this lens, reentry planning entails maintaining a configuration of dependencies. 
\citet{agapie2022longitudinal} found that goals were mediated through existing therapeutic relationships. 
In reentry, goals also rest on fragmented institutional arrangements, such as housing authorities, treatment programs, and community supervision, each imposing its own requirements, timelines, and eligibility conditions.
Along with the routines and relationships a person relies on, these arrangements form what we call the infrastructure of a goal.
We use \emph{infrastructure} in the relational sense \cite{star1996steps}, referring to what a practice runs on and what typically remains in the background until it breaks down.

HCI research has begun to engage with incarceration and reentry as sites of design and of inquiry into the institutional systems people must navigate \cite{gautam_enhancing_2024, ogburu2019towards, verbaan2018potentials, seo2021informal, grierson_design_2022, ogbonnaya-ogburu_supporting_2024}.
This scholarship shows that digital access and skills are necessary but insufficient. People can learn to, say, apply for a job online yet still be unable to sustain the conditions that make employment possible, such as stable housing \cite{Reisdorf_2022, gautam_enhancing_2024}.
Community members navigating reentry often have to revise their plans to keep fragile arrangements viable as institutional requirements, support relationships, and everyday stability shift.  
We argue that such revision is maintenance work.
It does not signal that the original plan has failed; it is a central practice through which plans remain meaningful under institutional fragmentation.

In this work, we partnered with West End P.O.W.E.R., a nonprofit supporting community members navigating reentry from incarceration, recovery, or both. 
Building on the organization's ongoing work, we co-designed FutureBound, a digital future-envisioning tool that preserves the expressive strengths of vision boards and adds structure for linking goals to resources, constraints, and dependencies. 
We introduced the tool through the organization's 12-week support program.
Our study draws on three workshops with 18 participants and follow-up interviews with 10 of them. 
We seek to address the following two questions:

\begin{itemize}
\item {\textbf{RQ1:} How do community members navigating reentry envision and articulate possible futures, including the assets, constraints, and dependencies that shape their goals?}

\item {\textbf{RQ2:} How do facilitated physical and digital future-envisioning activities support participants' sensemaking, reflection, agency, and revision of future possibilities under conditions of instability?}

\end{itemize}

Through this work, we make three contributions that build on each other. 
First, we offer a conceptual framing of \emph{revisability as maintenance work}. 
Drawing on repair-centered perspectives \cite{jackson2014repair, houston2016values}, we reposition goal revision as a legitimate form of agency, one that keeps plans viable when stability cannot be assumed.
Second, we provide an empirical account of \emph{dependency-aware future envisioning} in reentry and recovery.
We show how participants surface the fragile configurations of personal conditions, institutional requirements, and relational supports on which their goals depended.
Third, we translate these insights into design implications for future-envisioning tools embedded in community organizations.
We argue that such tools should make maintenance visible, normalize revision, and support mediated use across the care ecologies that shape which futures become possible.

%% file: litReviewUpdated.tex
\section{Literature Review}
\label{related-work2}

\subsection{{Reentry as Coordination and Maintenance Work}}

Reentry after incarceration, recovery, or both requires people to rebuild stability across fragmented institutional systems while carrying overlapping structural burdens \cite{halushka2019runaround, Petersilia_2001, gautam_enhancing_2024}. 
HCI scholarship on structural marginalization argues against treating marginalization as a stable group identity, attending instead to how institutional access, relational support, and material conditions shape what is imaginable and actionable \cite{erete2023method, OgbonnayaOgburu2020CriticalRT, wong2020culture, wong2021reflections, erete2018intersectional}.  
For people navigating reentry, these conditions shape access to services and technology, and also which futures feel possible and sustainable.

HCI has increasingly focused on designing \emph{with} communities whose trajectories involve navigating institutions and rebuilding stability across fragmented systems \cite{erete2018intersectional, liang2023surfacing, gautam2018participatory}.
Beyond basic needs, reentry requires emotional, social, and infrastructural support, which existing systems often fail to provide \cite{phelan2025social, gautam_enhancing_2024, stanojlovic2021targeting}.
Community members must stitch together brittle plans across agencies that do not see contingency and interdependence \cite{visher2008employment, halushka2019runaround, duffy2013recovery}, performing invisible articulation work to keep systems aligned \cite{strauss1988articulation, crabtree2006practical, suchman1996supporting, star1999layers}.
Digitized welfare services further intensify this work by embedding opaque and shifting eligibility rules into service delivery \cite{eubanks2018automating, Reisdorf_2021}.

Research on automation in public-facing services shows how technologies redistribute these burdens. 
In child welfare, algorithmic decision-support creates repair work when standardized systems overlook practice complexity \cite{saxena2021framework}. 
In homelessness services, frontline workers use discretion and workarounds to mediate between rigid prioritization systems and clients' situated realities \cite{tracey2024intermediation, tracey2024after}. 
Reentry organizations similarly mediate fragmented resources.
The scholarship cautions against technologies that obscure the relational judgment and care labor on which coordination depends \cite{rudnik2023care, star1999layers}.

HCI has responded to these conditions through participatory design, digital literacy interventions, and development attentive to the institutional settings in which tools are used.
Across this work, digital access and skills remain insufficient without trust-based learning environments \cite{seo2021informal}, integration with organizational workflows \cite{gautam_enhancing_2024, ogburu2019towards}, and alignment with broader goals such as employment and housing \cite{ogbonnaya-ogburu_supporting_2024, grierson_design_2022}.
These interventions address critical near-term needs, but less understood is how planning and future-envisioning function when futures remain contingent on shifting institutional pathways.

This gap directs attention to the coordination work of planning itself. 
CSCW scholarship on articulation work and infrastructuring shows that sustaining trajectories requires ongoing alignment among people, tools, and institutions \cite{star1999layers, karasti2014infrastructuring, le2012participation}.
Scholarship on ecologies of care locates this labor across individuals, peers, and organizations under precarity \cite{kaziunas2019precarious}.
\citet{steinhardt2014reconciling} frame planning as temporal alignment that bridges discordant rhythms, analyzing it both as plans are formed (`plans-in-the-making'') and as they are enacted (`plans-in-action''). 
In reentry, personal rhythms of healing and learning must align with program timelines, eligibility windows, and probation schedules. 
Planning thus becomes coordination across misaligned and only partially controllable temporalities. As uncertainty is generative, politically structured, and affectively lived \cite{soden2022modes}, revision and recalibration become central for people navigating reentry to keep futures possible. 

\subsection{Designing for Goal Setting and Resource Access}

In participatory design (PD) and speculative design, future envisioning can make sociotechnical configurations legible, enabling individuals to anticipate consequences, surface harms, and imagine alternatives before technologies are built \cite{harrington2019deconstructing}. 
However, future-oriented workshops may drift toward ``blue-sky ideation'' that overlooks constraints faced by marginalized communities \cite{harrington2019deconstructing}. 
In reentry, imagining a desired future cannot be separated from the resources, institutional pathways, and relationships through which it might become possible.

Goals are better supported when they are grounded in everyday circumstances and allowed to evolve \cite{ekhtiar_goals_2023}. 
Aspirations are similarly embedded in social and material realities, where what feels possible is shaped by institutions, relationships, and sociocultural resources and constraints \cite{toyama2018needs, kumar2019aspirations}. 
A gap remains, though, in making imagined futures actionable. 
Even well-defined goals can become demotivating without self-efficacy, resources, and institutional support \cite{bandura2013role, locke2015goal}.
Existing goal-setting tools often treat goals as linear targets to decompose and track completion \cite{li2010stage, epstein2015lived}, even though goals are often dynamic, shift between levels of abstraction, and require constant revision \cite{niess2018supporting, agapie2022longitudinal}.
Our concern, however, is not only that goals change. 
We note that under instability, pursuing a future requires ongoing work of sustaining the very dependencies on which it rests, so that the \emph{same} aspiration \emph{remains possible} as conditions shift.

To understand this ongoing work, we build on scholarship on maintenance and repair. 
Jackson's ``broken world thinking'' centers ``breakdown, dissolution, and change, rather than innovation, development, or design'' as ``the key themes and problems'' for understanding technology and its value \cite[p. 222]{jackson2014repair}.
\citet{houston2016values} extend this to show how repair practices reflect community values and sustain material life.
We distinguish \emph{maintenance}, the ongoing upkeep that keeps something working, from \emph{repair}, the work of restoring it after breakdown, while acknowledging that the two blur in practice \cite{jackson2014repair}.

This work further connects to scholarship in infrastructure studies.
Infrastructures comprise the relational background --- the standards, routines, and systems --- that support a practice and typically become visible only upon breakdown. 
Because infrastructure is inherently relational, what functions as seamless substrate for one actor may constitute the daily work or an explicit obstacle of another \cite{star1996steps}. 
\citet{star1999sorting} describe the analytical practice of bringing these background arrangements into the fore as ``infrastructural inversion''. 
We draw on this perspective by treating the necessary background conditions upon which a goal depends as that goal's infrastructure.
Maintenance has informed studies of infrastructure \cite{steinhardt2014reconciling}, care \cite{de2017matters}, and technological systems \cite{houston2016values}. 
Closest to our concern, \citet{steinhardt2015anticipation} describe \emph{anticipation work}, the practices through which people cultivate expectations of the future, build pathways toward them, and adjust those expectations as circumstances change. 
Their account comes from large scientific collaborations, where the futures in question are shared and institutionally backed. 
We extend this attention to personal futures held by people with few such guarantees, where the conditions a future depends on are largely controlled by others and can lapse without notice.


Reflecting on these bodies of work, we argue that designing for goal setting must go beyond helping individuals clarify what they want. 
It must also attend to whether the surrounding resources, support, and expectations keep those goals viable, so that people are not set up for failure.
Accordingly, FutureBound supports participants in articulating goals while linking them to resources, constraints, and dependencies. 
We also advocate for treating goal revision as a situated form of agency that requires empirical inquiry.

%% file: methodology.tex
\section{Methodology}
\label{methodology}

This paper reports on a research study as part of an ongoing collaboration with our partner organization, West End P.O.W.E.R., in Pittsburgh, Pennsylvania (USA).
The study described here took place between May and November 2025 and examines how community members engaged with iteratively developed future-envisioning tools across three workshops and ten interviews. 

\subsection{Background and Community Partnership}

The partner organization was founded by a community leader with lived experience. 
The organization challenges stigmatizing narratives by framing participants' past experiences as sources of growth and strength. 
Its 12-week program supports job readiness, digital literacy, financial literacy, leadership development, community organizing, anger management, and record expungement. 
The program runs three cohorts a year, and participants in our study were community members already enrolled in it.

Our study was part of the organization's ongoing programming and evolved from physical vision-board activities to the iterative development and evaluation of a digital future-envisioning system. 
This setting allowed us to examine future envisioning within the organization's existing programs and relationships of support instead of a standalone technology intervention.
This aligns with our vision and desire to be partners on the side \cite{gautam_enhancing_2024}.

\subsection{Research Approach} 
We began our study after receiving IRB approval from our institution. 
We worked as facilitators in the organization's existing programming, which built our familiarity with its priorities and practices, the people it serves, and their support networks. 
This grounded our design in approaches that were already acceptable and meaningful within the community.
We used a two-phase approach to understand how digital future-envisioning tools might support community members navigating reentry in sensemaking, building agency, and planning for their futures. 
Insights from situated practice informed the iterative tool design across both phases.

West End P.O.W.E.R. staff advised on workshop timing, supported recruitment, and helped align activities with participants’ needs and the ongoing curriculum. 
We, the research team, were responsible for the study design, workshop materials and tool iterations, data collection, interview facilitation, and analysis.

\subsubsection{Participants}
\label{participants}
Participants varied in age, reentry timelines, and housing and employment situations (see Table \ref{participant-demographic}). 
All eighteen participants in our study were women, reflecting the composition of the partner organization cohorts during our study period. 
Consistent with a trauma-informed approach \cite{tseng2025mitigating, elliott2005trauma}, we did not ask participants to disclose specific aspects of their history, such as whether they were in recovery, formerly incarcerated, or both. 
We focused instead on participants' present circumstances and envisioned futures. 
We use ``community members navigating reentry'' throughout to reflect this framing while centering the shared experience of rebuilding stability and envisioning futures amid structural constraints.

\subsubsection{System Description}
\input{systemdescNew}

\subsubsection{Phase 1: Vision-Board Workshops} 
The first phase focused on facilitating three in-person, vision-board workshop cohorts. 
We worked with six participants per cohort in May, August, and November 2025 (N=18), moving iteratively from physical envisioning practices to a fully digital system.

Each workshop lasted two hours and followed the same structure, an introduction to future-envisioning practices, a hands-on envisioning activity, and a post-session reflection. 
The exact duration varied across workshop cycles, but approximately 30 minutes were devoted to introductions and reflection prompts, 60 minutes to the hands-on activity, 20 minutes to post-session reflection, and 10 minutes to a break. 
Activities evolved across cycles as we moved from physical vision boards to direct interaction with the digital system.

Vision-boarding was already legible within the organization's emphasis on future planning and offered a low-barrier way to envision futures. 
Beginning with a physical, collaborative activity let us understand how participants externalized goals and made sense of their futures before embedding these practices in a digital system. 

\paragraph{May Workshop:} 
The first workshop focused on understanding participants' existing approaches to imagining the future. 
It emphasized building relationships and observing how envisioning practices aligned with existing community norms. 
The session began with a brief presentation on goal-setting, long-term planning, and future envisioning.
We created a printed booklet with reflective prompts across domains relevant to reentry (employment, housing, health, relationships, personal growth), which participants used as a starting point for their physical vision boards.
They reflected on their circumstances, strengths, needs, and goals by cutting and pasting magazine images to represent possible futures.

\paragraph{August Workshop:} 
The second workshop retained the paper-based vision board format, allowing us to refine prompts and facilitation based on insights from the first workshop. 
Towards the end of the workshop, we demonstrated the digital future-envisioning tool to participants. The participants did not interact with the system.

\paragraph{November Workshop:} 
The third workshop shifted to hands-on digital future envisioning. 
Participants used the digital tool to create vision boards. 
The system was designed to reflect practices participants had valued in the earlier workshops.
For example, participants could scroll through digital magazine pages in the tool and select images to add to their vision board, mirroring how they had perused physical magazines earlier.

Participants were then invited to present their vision boards and share stories about their envisioned futures with the group. 
Each workshop concluded with a post-session reflection on their experience, emerging insights, and whether the activity shifted how they viewed their future. 

\subsubsection{Phase 2: Semi-Structured Interviews and Digital Tool Evaluation} 
The second phase examined how participants engaged with the digital future-envisioning system. 
We used purposive sampling to recruit ten participants from the Phase 1 workshops, two from May, three from August, and five from November.
This let us capture variation in reentry timelines and prior exposure to the tool, and understand how familiarity shaped engagement.
We planned each session for 55--60 minutes, with approximately 15 minutes for rapport-building and an initial interview,
20--25 minutes for introducing the system and a guided user-probe activity, and 20 minutes for a post-session reflection interview.
The first part of the interview elicited participants' understandings of identity, agency, and the challenges they faced during reentry.

We then introduced participants to our digital tool, which supported them to create a vision board, upload and add images, write notes, and map connections between their current state, intermediate goals, and long-term aspirations (see Figure \ref{fig:futureEnvisioningTool}). 
We followed a think-aloud protocol \cite{lewis1993task}, in which participants explored the tool while narrating their thoughts, decisions, and emotional responses. 

After participants explored the tool, we resumed the interview, focusing on any shifts in how they perceived their goals or future possibilities, whether the visual mapping supported sensemaking, agency, or reflection, how useful they found connecting goals to real resources, and what they wanted improved. 
Our evaluation of the digital tool was session-based. 
Participants interacted with the system during the workshop or interview, or both. 
We did not evaluate continued use of the digital tool over time.

\subsection{Data Collection and Analysis}

We collected data across both phases of the study through field notes, workshop booklets, interviews, think-aloud sessions, and follow-up reflections. The booklet responses captured participants' self-described strengths, support systems, and needs.  
Workshop field notes documented participants' interactions with envisioning activities, moments of difficulty or engagement, and recurring responses that informed subsequent design iterations. 
We audio-recorded and transcribed the interviews, which yielded over seven hours of recordings across the ten participants. 
Individual sessions lasted approximately 42 minutes on average.

We analyzed the data using reflexive thematic analysis \cite{braun2006using, braun_reflecting_2019}. Two researchers first reviewed transcripts and conducted inductive coding to generate 556 initial codes, which were iteratively consolidated through collaborative discussion into 18 higher-level codes.
These captured reentry experiences and interactions with envisioning practices and digital tools. 
We developed these higher-level codes (e.g., ``housing and employment concerns,'' ``skills as personal strengths,'' ``vision board impact'') into preliminary themes and refined them into the three overarching themes that structure our findings.

\subsection{Researcher Positionality}
Our research team consists of a PhD student in Computer Science, an undergraduate researcher, and a faculty member with over ten years of community-based design experience, all affiliated with a large public university in the region.
Two of us are international researchers, and one is American. 
Neither we nor our immediate family members have been directly affected by the criminal justice system, which makes us outsiders in many respects to the community with whom we conducted this research. 

We acknowledge that our outsider status shaped what we could see, ask, and design for. 
We also brought assumptions about planning and future envisioning that may not reflect how participants navigate precarity or define progress in their lives.
We relied on the organization's knowledge of the context and the participants, built activities around their ongoing work, and worked with them to frame the tool and the pace of the workshop. 
We also sought to ensure that staff could redirect decisions when they conflicted with participants' needs, for example, around image representation, as discussed in Section ~\ref{implications}.

This reflects our priority of building a reciprocal academic-community partnership.
Through our ongoing collaboration with the organization, we are learning what it means to be ``a partner on the side'' \cite{gautam_enhancing_2024}: to follow the lead of those with lived expertise and to design \emph{with} them to support their ongoing efforts.

We approach this work from a commitment to justice and fairness. 
The structural realities facing people returning to their communities often make envisioning a stable future, let alone achieving one, difficult \cite{halushka2019runaround, Petersilia_2001}.  
We see our work as contributing, in a tiny way, to a more just world, one in which tools designed to support people account for these structural constraints.
We do not claim to fully understand the experiences of community members navigating reentry.
Instead, this paper presents our attempt to center community members' knowledge and aspirations in the design process.

\subsection{Ethical Considerations}

Given the vulnerability of our participant population, we took several steps beyond standard IRB protocols to ensure ethical engagement. 
Our partner organization staff reviewed our research design and provided feedback on culturally appropriate facilitation. 
Our informed consent process emphasized that participation was voluntary and would not affect participants' standing in the organization's program.
As noted in Section~\ref{participants}, we did not ask participants to disclose their history. 
This decision limits our ability to disaggregate findings by specific population. 
Staff members were present during the workshop to respond to sensitive disclosures or any other difficult circumstances.
We conducted interviews at times and places convenient for participants, including fast food restaurants such as Wendy's and McDonald's, and online on Zoom.
Interview participants received \$50 as compensation, a decision made in consultation with the organization to ensure reciprocity without creating coercive incentives.

\begin{table*}[ht]
\centering
\begin{tabular}{l c c c c}
\hline
\textbf{Participant ID} & \textbf{Age Range} & \textbf{Time elapsed since reentry}  &
\textbf{Participated in Study} \\
\hline
W1P1 & 25-34   & 2+ years & Workshop \\
W1P2 & 45-54   & 2+ years & Workshop and Interview \\
W1P3 & 55-64   & N/A &  Workshop \\
W1P4 & 35-44   & 6 months - 1 year &  Workshop and Interview \\
W1P5 & 55-64   & 1 - 6 months &  Workshop \\
W1P6 & 55-64   & 6 months - 1 year & Workshop \\
W2P1 & 45-54   & N/A &  Workshop \\
W2P2 & 55-64   & 1 - 6 months & Workshop \\
W2P3 & 35-44   & 1 - 6 months &  Workshop and Interview \\
W2P4 & 45-54   & 1 - 6 months & Workshop \\
W2P5 & 55-64   & 1 - 6 months &  Workshop and Interview \\
W2P6 & 35-44   & 1 - 6 months &  Workshop and Interview \\
W3P1 & 35-44   & 1 - 6 months &  Workshop and Interview \\
W3P2 & 35-44   & 6 months - 1 year &  Workshop and Interview \\
W3P3 & 35-44   & 6 months - 1 year & Workshop \\
W3P4 & 25-34   & 6 months - 1 year &  Workshop and Interview \\
W3P5 & 35-44   & 6 months - 1 year & Workshop and Interview \\
W3P6 & 35-44   & 6 months - 1 year & Workshop and Interview \\
\hline
\end{tabular}
\caption{Participants in the two phases of the study. W1, W2, and W3 correspond to May, August, and November workshops, respectively. `N/A' for time since reentry indicates participants did not disclose this information, or reentry was not relevant for them.}
\label{participant-demographic}
\end{table*}

%% file: systemdescNew.tex
\begin{figure*}
    \centering
    \includegraphics[width=1\linewidth]{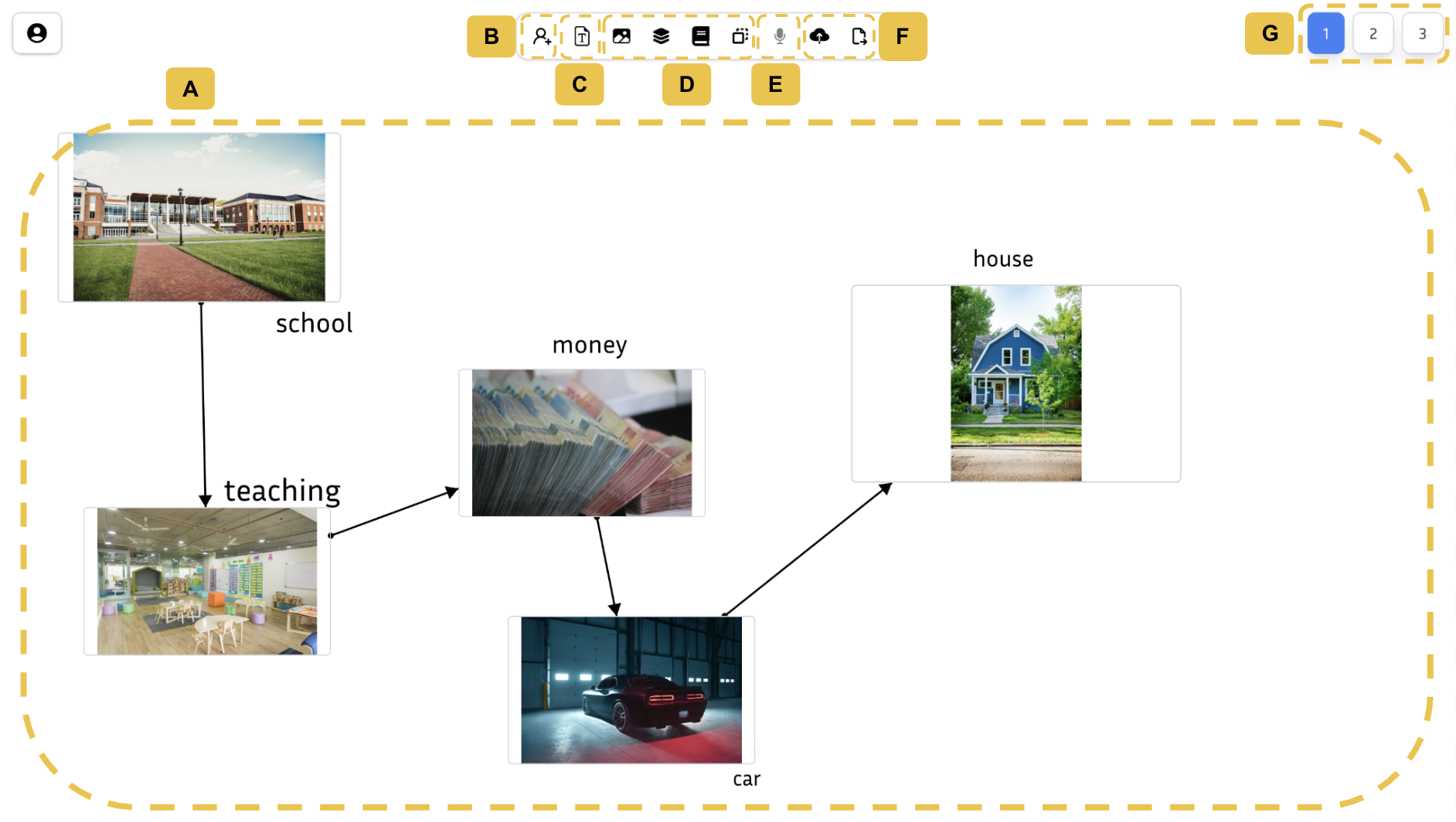}
    \caption{An example screenshot of FutureBound, which helps users  create goals and the actionable steps to achieve them.}
    \label{fig:futureEnvisioningTool}
\end{figure*}

{FutureBound is a web-based application designed for desktop or laptop browsers and does not currently support mobile use. 
Its design builds on the physical vision-board practices participants already valued. 
It translates the acts of browsing magazines, cutting out images, and arranging them on a board into digital equivalents, easing access to imagery without requiring magazines or open-ended internet searches. 
Unlike general-purpose image curation platforms such as Pinterest, FutureBound is organized around planning.
It structures envisioning across a current state, a pathway, and a desired future, supports linking goals to the conditions and resources they depend on, and records the narrative behind a board through audio storytelling.
}
We implemented the front-end interface in React and integrated it with the Unsplash API\footnote{\url{https://unsplash.com/developers}} for image retrieval, along with backend services for image recognition, story recording, user authentication, and cloud storage.

The system supports parallel goal setting and planning by providing three vision boards (1G). 
Each board has a central canvas area (1A) where users can visually express ideas, feelings, and goals. 
Specialized tools allow users to add mentor figures (1B), text annotations (1C), and images sourced from Unsplash (1D).
Users can also extract images from photographs of their prior physical vision boards, find images in imported digital magazines, or use the image search tool (1D). 
These options were intended to lower barriers for people with varied digital literacy.
The Storytelling tool (1E) allows users to record audio narration explaining the thinking behind their vision board. 
Users can also create connections between elements using arrows and lines.

To support offline use and protect user privacy, the system supports users to download their vision boards as JSON files (1F) and later upload them to revisit and revise their boards. 
Logged-in users may alternatively save boards directly to their accounts. 

%% file: findings.tex
\section{Findings}
\label{findings}



We present findings from the ten follow-up interviews examining how community members navigating reentry engaged with the future envisioning tool and activity. 
Although participants came to the interviews with varying prior exposure to the system (W1: physical-board workshop only; W2: physical-board workshop and demo; W3: workshop with hands-on use of the tool), all interviewees engaged with the digital tool during the semi-structured interview. 
The workshop data consisted of researcher field notes capturing how participants interacted with the materials, and facilitators and peers.
These notes were observational and did not capture speech verbatim.  
We do not present them as quotations. 
Instead, they contextualize participants' engagement and how workshop practices shaped later iterations of the tool.

All participants found the digital vision board accessible after initial orientation. 
Some found it intuitive; W3P2 noted that \textit{``navigation was the best part''.}
Others faced a steeper learning curve. 
W3P5 remarked, \textit{``It was all unfamiliar. Yeah, technology's all new.''}
The facilitated workshop helped bridge these gaps as W2P6 reflected, \textit{``that was actually pretty easy with you, ... how you showed me and talked to me.''}
As participants grew comfortable, they described a sense of ownership over the process. 
W2P5 observed, \textit{``I think if you learn how to work this around ... when you do it by yourself, I think that makes you like more powerful.''}

We developed three interconnected themes. 
The first two concern the structure of goals in reentry contexts. 
They show that goals depend on fragile configurations  of personal conditions, relationships, and institutional requirements.
First, participants described recovery as an ongoing process requiring continuous maintenance and used the vision board to map what needed to be sustained. 
Second, the envisioning activity functioned as a form of asset accounting, helping participants recognize existing strengths while surfacing gaps and dependencies that planning would need to address. 
The third theme concerns keeping those configurations viable over time. 
Participants treated goal setting as an agentic practice through which they articulated aspirations, confronted constraints, and revised trajectories as conditions shifted.
These themes point to two aspects of planning under institutional fragmentation --  goals are dependency-laden (what goals depend on), and they are maintained through revision (what it takes to keep them viable).
As W1P4 put it, \textit{``goals [can] change like tomorrow. Goals, you can just update it as you go.''}

\subsection{Reentry and Recovery as Maintenance Work}
Participants described their recovery as an ongoing journey rather than a milestone to arrive at. 
Across workshops and interviews, recovery was narrated through time, routines, and the institutions and relationships participants relied on. 
Progress depended on committing to sobriety, managing mental health, and securing everyday stability, such as housing and work.
The vision board prompted participants to articulate what going forward would look like, often by naming conditions they needed to sustain and reflecting on how easily those conditions could shift. 
In this way, the tool helped participants map trajectories of maintenance and resilience more than it helped them mark achievements.

\subsubsection{Mapping What Must Be Sustained}
Many participants envisioned their futures in relation to sobriety timelines that were concrete but cautious. 
They spoke of their time in sobriety with pride and with vigilance, treating it as an achievement that required ongoing work. 
W1P2, reflecting on rebuilding clean time after a relapse, shared, \textit{``Once I lost my seven years ... I'm coming up on five years clean in November. So I'm sitting still. I want to write my book. I want to find a place I can call home with the job. For my relationship with my higher power to get stronger.''} 
Her account weaves sobriety into a broader constellation of goals, such as housing, work, and spirituality, all of which require sustained attention. 

For several participants, the possibility of relapse remained present in the background of planning.
They placed sobriety at the center of their recovery while recognizing it could disrupt everything else. 
As W3P1 noted, \textit{``I'm an addict, and so, ... that's always a big barrier, and that could literally throw any of these dreams or goals, ... right off course.''}
Accounts like these suggest that future envisioning did not erase uncertainty but did help clarify priorities. 
For example, W2P6 explained, \textit{``Right now, I'm okay with not working and all that, and doing the disability, because I can put my recovery first. Because anything I put in front of recovery --- my dog, my mom, my girlfriend, whatever it is --- that'll be the first thing I lose if I go back to it} [relapse].''
The vision board activity supported participants to represent this sequence of priorities, with recovery at the center. 

Managing mental health was at the center too.
Participants emphasized that maintaining sobriety required attending to emotional regulation, anxiety, and long-standing mental health conditions. 
W2P3 explained, \textit{``My health, my mental health is very important to me, because if my mental health isn't right, then I can't be right.''} 
Participants described replacing maladaptive coping strategies with routines they felt were more stabilizing.
For example, W1P4, while sharing her experience with dealing with depression, said, \textit{``I start not focusing more on like social media ... So I just like stopped going on it. And then I got a coloring book ... Especially with the nice weather. I try to take advantage of it and go out.''}

These practices appeared directly on participants' vision boards as commitments they wanted to sustain.
W3P1 reflected, \textit{``I remember I put a lot of, like, yoga and meditation, into mine, because that's a big part of my life. With vision boards, you're kind of ... manifesting ... by seeing it every day ... I actually do wanna work on it.''} 
By placing ongoing practices alongside future goals, participants used the boards to articulate what maintenance would require, making visible the continuous work that recovery demands.

\subsubsection{Naming Fragile Conditions}
\label{sec:fragileConditions}
Participants consistently described everyday stability, particularly housing and employment, as both essential to their recovery and difficult to attain. 
Images of houses, cars, financial and physical wellness, and daily anchors appeared repeatedly on their vision boards.
These images suggest a focus on the here and now more than on a distant future.
W3P4 explained how the vision board activity helped in narrowing broad desires into attainable goals, \textit{``I wanted to secure some form of employment as well as stable housing that I can rely on. As well as a possibility of finding a life partner ... It was very helpful to actually narrow down the scope of the ideas in my head to real, realistic, attainable goals as well as providing smaller steps that I could take to actually accomplish said goals.''} 

At the same time, participants described their sense of stability as fragile, especially while transitioning out of halfway houses or temporary housing facilities. 
Participants described housing uncertainty as the most persistent challenge to a stable future. 
W3P4 shared, \textit{``I'm no longer in the halfway house I was in. I'm not in my own place, but I'm in transitional housing right now, working towards being able to get my own apartment.''} 
Even when participants were employed, structural barriers made securing housing difficult. 
W3P1 described the lack of communication from her housing authority, \textit{``It was really hard to get in touch with the housing authority downtown, and I don't even think I ever met my case manager within the eight years I was there. And it's really difficult too, because you have to prove that you work, you have to have a good credit score, and it just kind of defeats the purpose of having accessible housing for low income.''}

Similar tensions surrounded employment. 
While participants viewed their work as essential for financial stability, several described keeping their current jobs as a strategy of endurance rather than a means of achieving stability and security. 
W1P2 echoed this sentiment, sharing, \textit{``I'm at a job that I really don't like, but I show up when I put 110\% every day, and I trust the process.''} 
Workplace inflexibility also compounded recovery challenges. 
For example, W2P6, managing multiple mental health conditions, explained, \textit{``I did work, and I get jobs here and there ... but I'm so high-strung -- my manic stages -- it's really hard to have a job ... And then my depression -- I will sleep for days, I will not shower, I will not eat, manic, I won't sleep -- my mind don't stop, my body don't stop, like, it's been like this my whole life.''} 

In these accounts, we hear both the necessity of stability and the fragility of the conditions that enable it. 
Housing depends on employment, employment is strained by mental health, and mental health is threatened by housing precarity. 
The vision board activity helped participants name these interdependencies. 
The tool's connecting features --- arrows and lines between elements --- helped them visually trace dependency chains, making the relationships between conditions explicit.
Participants did not see recovery as a linear path toward fixed goals, and the envisioning activity made visible what had to be maintained, what remained precarious, and why plans had to stay open to revision as conditions shifted.

\subsection{Identifying Assets and Needs through Future Envisioning}

Moving beyond current challenges, participants reflected on what would be required to make their envisioned futures possible. 
While building their vision boards, they identified and evaluated resources they could draw on, gaps they would need to address, and dependencies on people and institutions that they would need to engage. 
The activity also surfaced how fragile these assets were, since they needed to be maintained and often depended on others' cooperation.

\subsubsection{Recognizing Strengths as Actionable Resources}
The future envisioning activity created space for participants to reflect on their existing personal strengths as assets relevant to their reentry journey. 
Participants described resilience, discipline, caregiving, optimism, and perseverance as capabilities they rely on to navigate recovery and everyday stability. 
W1P2 described this reflective capacity, noting her ability to \textit{``sit still today, reflect back on my day, see what I might have done wrong, what I could get better at, and what I did right. I wake up with a peace of mind. I wake up with a desire to live. I have a hope shot.''} 

Importantly, these strengths were not abstract traits.
Participants connected them to desired identities, roles, and responsibilities, linking who they were to who they hoped to become.
Caregiving was a particularly salient example. 
W1P4, drawing on 14 years of experience in nursing, explained, \textit{``I think that's why I've been in the nursing field for so long, because I really like taking care of people. About 5 years from now, I want to ... open my own nursing home.''}

Similarly, participants who described themselves as disciplined and resilient connected their qualities to goals such as sustained sobriety, stable employment, and returning to school. 
W3P4 emphasized optimism as a capability for navigating long-term recovery and daily hardship, noting, \textit{``Some personal skills in particular that I rely on daily are the unwavering optimism. Resilience, because even when things got bad, I was able to stand back up again ... it helps me shave off the bad days, so I'm able to keep going forward with everything.''} 

Participants also recognized family members, sponsors, therapists, peers in recovery, and program staff, as critical relational resources. 
Family served as both motivation and a source of accountability, often described as reasons for sustaining sobriety and pursuing stability. W3P5 reflected on the role of family in shaping her near-term plans and longer-term aspirations, \textit{``I'm almost 8 months sober, so that's a big achievement. I'm moving to [a nearby town] to be with my kids. It's been a rough journey. But now, since Colorful Backgrounds [our partner organization's program], with having a computer, I want to go back to school, and I'm planning on moving to [a nearby town] with my children and go to school out there.''}  
Envisioning futures involved making relationships visible and, in some cases, positioning repair and preservation of relationships as part of the plan.

Support networks extended beyond family. 
For instance, W2P5 described maintaining an extensive, distributed support network through meetings and introductions, \textit{``My girls, they are my support, going to meetings, my sponsor, stuff like that ... They introduce you to other people. You just meet different people. And you get support through them ...''} 
      
At the same time, participants emphasized the fragility of these relational supports. 
Explaining the impact of losing both a sponsor and a mental health counselor, W2P6 shared, \textit{``My sponsor passed away October last year. That woman knows me better than my own mother, and I have a very good relationship with my mother, so I lost her. And then in December, my mental health counselor retired. And I had her for 2 years, so it's like I already went into that relationship, having her as my therapist. ... It's hard for me to do this again because of all this.''} 
Future envisioning did not erase these challenges, but it helped participants recognize their support network as an asset that required ongoing maintenance.

\subsubsection{Surfacing Gaps and Institutional Dependencies}
\label{sec:gapsInstitutionalDepdencies}
The envisioning activity also surfaced resources that were missing, mismatched, or difficult to access.  
Participants positioned institutional resources such as recovery programs, housing services, employment assistance, and educational opportunities as gateways to their envisioned futures. 
W3P4 recounted how a halfway house provided necessary support during reentry, \textit{``One of the biggest supports that I received towards being able to start actually living my life instead of just surviving was [a] halfway house that I was in. That gave me the tools, resources, and support necessary to take steps that I wouldn't normally take without a fallback plan.''} 
Conversations about desired futures also revealed educational aspirations and the resources needed to realize them.  
W3P5 noted that \textit{``grants for schools [are] going to be a big help.''} 
W3P6 described programs combining financial support with employment pathways, sharing, \textit{``there's actually a program that will pay you to go to school and everything and then they also give you a job where, like nursing home as well, so I'm definitely gonna get back involved with that.''} 

However, participants also described institutional resources as difficult to access and navigate. 
W3P5 reflected on searching for assistance online and feeling overwhelmed by the information, \textit{``And like if I type in, I don't know, money management. There's 80 things, you never know what to do, what's a scam. So that's frustrating.''} 
Others described misalignment between available resources and their personal circumstances. 
W1P3 shared, \textit{``There's a lot of things for like babies and stuff. I don't have any children.''}  

This friction was particularly salient when participants tried translating institutional options into actionable steps.
Even when they recognized potential services, they struggled to determine what to do next, in what order, or whether they were eligible. 
W2P3 reflected on how the envisioning activity prompted this kind of assessment, sharing, \textit{``It made me think about both [assets and challenges], what I am equipped with now, what I have to work with now, and what I will need to work with in order to get to where I want to be.''} 
The vision board did not resolve these systemic barriers, but it helped participants identify existing resources, what remained uncertain, and where gaps in institutional navigation could derail otherwise meaningful plans. 

\subsection{Sustaining Agency Through Revisable Goal-Setting}

The previous two themes examined what goals depend on, that is, the fragile configurations of conditions, relationships, and institutional requirements that make futures possible. This theme turns to what it takes to keep those configurations viable over time.
Participants did not always treat their goals as endpoints. 
Instead, they described goal setting as an ongoing process of prioritizing what mattered now, identifying actionable pathways, and anticipating constraints, and through this, gaining some control over aspects of their futures. 
We noticed that this agency was closely shaped by the support participants could draw on, including staff, programs, and relationships, and by their awareness of institutional resources. 

\subsubsection{Articulating Goals from Abstract Hopes}
Participants often started with abstract hopes such as \textit{``doing better''}, \textit{``having stability''}, \textit{``getting my life together''}.  
These visions were hazy and loosely specified.
Through the vision board activity, they translated these abstract desires into named goals across domains such as housing, education, employment, family, and health. 
Selecting, labeling, and placing specific goals on the board externalized their aspirations into explicit commitments that the participants could reason about, compare, and prioritize.

For many, this visualization shifted goals from thoughts to concrete commitments, as heard in W3P1's account, \textit{``to see it in front of your face instead of just thinking about it really makes it feel more real. It starts to turn it into less of a dream and more of a goal.''} 
W3P6 echoed this, explaining how the activity built her confidence that her goals were possible, \textit{``Making that vision board, ... really gave me hope, ... it has given me, like, I can do this, you know, ... these things are obtainable, ... I'd actually be able to get this house now that I'm clean and doing the right thing. And not that I didn't believe that before, but I guess doing that vision board had really made it come true. Like even just in my mind, ... you know you can do something, but when you're putting something on paper, it makes it obtainable to you, like, more real.''} 
The digital tool's image search function supported this articulation process.
W3P2 described how searching for images helped her externalize internal states, \textit{``The part where you search the images, and, you know, put what's in your mind. Like, something that you want.''}

When participants named specific goals, such as a house, a job, or credentials, they confronted questions of feasibility, sequencing, and dependency.
W3P4 reflected on this challenge, explaining, \textit{``The only thing challenging for me in that regard was what order I could actually feasibly accomplish things in. Because I've always had an idea of trying to do things backwards without realizing that that was not actually attainable.''} 
These reflections helped participants recognize what they could realistically pursue, given practical constraints and questions of sequence.

\subsubsection{Making Progress Legible}
Beyond long-term visions, identifying near-term actions and intermediate milestones seemed important for sustaining momentum. 
During the workshops, we nudged participants to think through intermediate steps rather than jumping directly from aspiration to outcome. 
W2P3 explained how this led her to consider the process required to achieve her goals, \textit{``It made me put a lot of more things in a perspective of what I need to do in order to achieve what I want to achieve. Like, put different steps into play, what's required for me to get to those points.''} 

Having both short-term actions and long-term visions visible at once allowed participants to situate immediate steps within broader life trajectories. 
W3P4 described how this visibility supported her sense of progress, \textit{``I would say that you guys have created a very well-thought-out program. I feel a lot more confident in actually being able to see what little steps I can take towards accomplishing my bigger goals, and it's awesome to be able to actually track the progress I've made by being able to cross off steps along the way.''} 
Here, agency was supported by making progress legible in increments, a form of accounting that showed participants how far they had come while clarifying what remained. 

\subsubsection{Revising Trajectories to Keep Plans Viable}

Participants consistently framed their goals as iterative, reflecting the uncertainty of reentering into society and circumstances that can shift quickly. 
Revisability allowed participants to adapt their plans in response to new information, changing circumstances, or an evolving sense of their position. 
In this context, the capacity to revise, instead of a commitment to a fixed trajectory, functioned as an indication of their agency. 

Several participants described vision boards as artifacts that could be revisited or remade as priorities changed, and some reflected on their prior experiences of maintaining multiple physical boards or returning to earlier ones.
W1P2, for example, noted, \textit{``I have my vision board on my door, my bedroom door, one of many. I have three of them. Those different parts of the journey of my life and where I was with my thought process with my spiritual, mental and emotional process is to me that's very important. So I still reflect on what I put on my vision board.''} 
The digital format supported this iterative orientation.
W1P4 identified revisability as valuable, remarking, \textit{``What if I think it's something I want to do next week and go back to edit the Vision Board, like not erase it, you know, but just add more photos''}  adding, \textit{``I think that's what the best thing about the website that you created is, you can always go back.''} 
Unlike physical vision boards, where revision often means starting over, the digital tool allowed participants to adjust and extend their plans while preserving what came before.
The save-and-return functionality (downloading boards as JSON files or saving to accounts) meant that revision was an additive action. 

Revisability also enabled participants to reflect on their progress. 
W1P2 reflected while comparing earlier and current boards, \textit{``It’s nothing like the vision board I did when I was in treatment prior to going to the halfway house, it really shows me my growth. It shows me my vulnerability and my willingness.''}
W3P6 echoed this, noting, \textit{``it's more growth for me, like, it made me see where I was at today, and how far I've come.''}
The tool functioned as a record of change, making visible a trajectory that might otherwise remain implicit.

Importantly, participants connected revisability to a sense of control over their futures, which opened space to consider that their long-term aspirations might change. 
W1P4 acknowledged this openness when discussing a five-year horizon, \textit{``maybe in five years I don’t want the nursing home''}. 
In her account, we hear shifting of aspirations as a legitimate approach to successfully integrate into society.

Participants did not commit to a single, static future. 
Instead, they prioritized maintaining the capacity to revise their trajectories. 
FutureBound supported this as a revisable artifact that normalized adaptation, enabled reflection across time, and helped participants renegotiate goals over time without treating those changes as failures.

%% file: discussion.tex
\section{Discussion}
\label{discussion}

Our findings suggest that future envisioning can extend beyond imaginative projection and become practical, ongoing work under conditions of recovery and institutional fragmentation. 
Goals remain viable only when people can maintain the conditions and relationships that make them possible.
Participants used the vision board activity to externalize aspirations, identify what they possessed and lacked, surface interdependencies, and treat revision as a legitimate response to shifting circumstances. 
Future-envisioning tools in reentry can thus do more than support creativity.

We argue that revisability should be a core mechanism for sustaining agency under instability, and we theorize revision as maintenance work.
We introduce dependency-aware future envisioning as a framing that treats goals as contingent configurations of personal conditions, institutional requirements, time, and relational support, that is, as having an infrastructure of their own. 
These insights emerged within a facilitated workshop setting. 
The tool did not operate in isolation. 
Our facilitation shaped how participants engaged with the boards, interpreted what they saw, and worked through difficult reflections.
Building on these findings, we articulate design implications for future-envisioning tools that work within, and depend on, the organizational care ecologies around them.

\subsection{Future Envisioning as Maintenance Work}

Scholarship on maintenance and repair has largely concerned things that already exist, including infrastructures, devices, and material arrangements \cite{jackson2014repair, houston2016values}.
Our participants did both maintenance and repair.
For example, W3P1 kept yoga, meditation, and sobriety on her board so that she would be ``seeing it every day,'' a form of maintenance.
W1P2, rebuilding clean time after she ``lost [her] seven years,'' was doing repair.

We extend this orientation to futures. 
We build on Steinhardt and Jackson's account of anticipation work \cite{steinhardt2015anticipation}. 
They studied scientists who worked toward futures that their institutions had committed to building. 
Our participants maintained futures that institutions often overlooked, and the work of keeping those futures viable fell largely on them.
W1P4's remark that ``goals change like tomorrow'' and W1P2's description of keeping ``one of many'' boards across ``different parts of the journey'' point to the maintenance of possibilities rather than of any specific object.
This framing shifts attention toward the conditions that make a future viable, the labor required when they change, and how that labor is distributed across individuals, organizations, and institutions. 
For HCI and CSCW, this opens a set of questions --- What conditions sustain a desired future? What becomes fragile when they change? Who performs the work of renegotiation, and what support makes that work possible?
It also changes what counts as progress. 
Remaining housed or sustaining sobriety is active maintenance, not stagnation, as participants showed by placing recurring practices alongside their aspirations on the canvas. 

%

Revisability is not unique to reentry, but three factors make it especially important in this context.
First, the dependencies are often institutional and structural, not only matters of personal preference. Whereas material infrastructure typically announces itself upon breakdown \cite{star1996steps}, the institutional conditions of a future can fail quietly.
An eligibility window closes or a case file stalls, and nothing visibly breaks. 
But for the individual, a provisional future moves out of reach.
Maintaining a future therefore requires noticing, performing a proactive infrastructural inversion \cite{star1999sorting}, to surface which background conditions a goal rests on and whether they still hold before a pathway collapses.

Second, the stakes of failing to keep a plan viable are disproportionate for populations already at the margins.
Participants' dependencies were tightly coupled. 
For example, W3P1 described how a relapse could derail all of her goals at once.
Under such stakes, repair may change what is being repaired.
A repaired device still answers to what it was for, even when repair is inventive \cite{jackson2014breakdown}.
Repairing a future may mean revising the destination, as when W1P4 allowed that she might no longer want to open a nursing home in five years.
But the systems around people in reentry tend to read change as failure.
Success in reentry is commonly measured as the absence of recidivism \cite{weisberg2013meanings} and reentry programs often locate setbacks in individual responsibility \cite{miller2014devolving}.
Reading revision as maintenance asks us to consider whether the conditions for a future a person still wants remain in place.

Third, the capacity to do this maintenance work depends on support networks, organizational resources, and institutional flexibility that may or may not be available.
Participants carried much of the work of keeping their futures viable. 
But they were not alone. 
For example, W3P4 credited her halfway house with giving her a fallback plan that made new steps possible.
However, such support can lapse. 
When a housing authority does not respond (Section \ref{sec:fragileConditions}), the maintenance it defers does not disappear.
It only shifts to the person whose future depends on it. 
We posit that maintenance raises a question of distribution, of who is expected to keep a future viable and who is excused from that work.

FutureBound supported revisability, the capacity of a plan and its artifact to change while keeping their history, in modest ways.
Its digital format made revision additive.
Participants could layer new goals over earlier ones, keep boards in parallel, and compare past and current boards.
W1P4, for example, stated that she valued being able to go back.
The tool did not maintain anyone's future.
It offered a place where that work could be externalized and discussed.

\subsection{Dependency-Aware Future Envisioning}
\label{sec:dependency}


We use \emph{dependency-aware future envisioning} to describe goals as configurations of conditions on which their viability depends. 
Planning, in this lens, involves making those conditions visible and negotiable. 
The conditions participants surfaced were personal (e.g., recovery stability), relational (e.g., family ties or sponsor support), institutional (e.g., eligibility requirements), and sequential (what must happen before something else becomes possible). 

This attends to scholarship on the mismatch between social requirements and technical feasibility, where systems fail by simplifying the realities they support \citep{ackerman2000intellectual}.
In reentry, this gap becomes acute because people must coordinate across fragmented, surveillance-heavy, and often punitive systems while managing basic needs and recovery \citep{halushka2019runaround, riggs2015reflections}.

Dependency-aware envisioning builds on and extends two bodies of work.
First, aspirations-based design treats aspirations as situated and evolving  \cite{kumar2019aspirations, toyama2018needs}.
Our findings add that under institutional fragmentation, aspirations may be held provisionally as protection against conditions that may not hold. 
Their evolution is driven by the fragility of supporting conditions as much as by individual growth.
Second, assets-based design understands capacities as enacted through contexts and relationships \cite{wong2020culture}, while cautioning against using assets as a substitute for confronting structural disinvestment \cite{emejulu2022critical, hardy2024turn}.
While we concur with this, our findings complicate the assumption that assets are somewhat stable. 
In our framing, an asset is a dependency that is currently available.
W2P6's loss of both a sponsor and a mental health counselor shows that relational assets also require maintenance. 
Dependency-aware future envisioning therefore centers temporality, treating aspirations and needs as continuously evolving and assets as available now but not necessarily later.

This framing also changes the questions asked in goal setting and planning.
Instead of asking whether a goal is clearly stated, we believe we should ask what conditions sustain it, how fragile each one is, and who can act on it.
This separates conditions a person can act on, such as a daily routine, from conditions they have little control over, such as eligibility rules or the availability of affordable housing.
W3P1's account of being unable to reach her housing authority over eight years illustrates the latter.
Articulating the separation clarifies the boundaries of individual agency and where support from staff, organizations, or institutions may be necessary.

FutureBound made these configurations inspectable through its three-part structure of current state, pathway, and desired future, along with connecting arrows that let participants and facilitators see where a pathway rested on a condition that might not hold. 
W2P3's reflection on ``what I am equipped with now ... and what I will need to work with'' exemplifies dependency-aware planning with assets, gaps, and next steps on a single canvas.

Making dependencies visible also has value beyond individual workshops. 
Participants across cohorts identified similar dependencies, including housing authorities that did not respond, resources mismatched to their circumstances, and eligibility rules that restricted their possibilities. 
These patterns forms a cumulative account of where institutions are failing. 
Such artifacts could serve as evidence that organizations use to advocate against the structural conditions that generate instability, moving design and research from individual toward structural engagement \cite{tracey2024after, dombrowski2016social}. 
We did not study this advocacy use and see it as a direction for future work. 
We believe, however, that it distinguishes our approach from the ``blue-sky ideation'' \cite{harrington2019deconstructing} that future-oriented workshops often drift toward.

\input{implications}

\subsection{Limitations and Future Work}

Our study has several limitations that inform directions for future work. 
Our findings are situated within a specific organizational and community context and do not capture the range of people in recovery and reentry.
All of our study participants were women, which does not reflect the broader U.S. reentry population.
This composition may have shaped the dependencies participants surfaced and how they engaged with the tool. 
This seemed particularly evident in discussions of caregiving, housing, recovery, and family relationships.
Although we do not claim these patterns are unique to women, our sample limits our ability to assess how they differ across gender-diverse reentry trajectories. 
Future work should examine how dependency-aware planning and revisability as maintenance play out across these trajectories.

Our evaluation was also limited to sessions.
Participants used the tool during workshops and interviews, and our claims about maintenance over time draw on participants' accounts of their vision-board practices and the tool's affordances for revision.
We did not observe use across an extended deployment. 
We believe that longitudinal studies should examine individual and staff-mediated use over time, particularly how revisable boards support adaptation across evolving reentry trajectories.

Further, we noticed differences in engagement tied to digital literacy. 
Some participants navigated the system independently, while others required substantial guidance. 
Although facilitation helped mitigate barriers, future work should examine how digital literacy intersects with future-envisioning practices and how tools can better scaffold participation without burdening the staff or relying on researcher facilitation.
Future iterations could provide in-context guidance through feature tutorials, introduce tool complexity gradually, and support peer assistance among participants with different levels of digital familiarity.

Our position as university researchers also shaped the partnership and participants' engagement. 
Despite efforts to build trust over time and embed our work within existing organizational relationships, power differentials remained.
Participants may have presented themselves or their experiences in ways shaped by our presence.

Future work can build on these limitations.
Integrating future-envisioning tools with resource navigation systems, such as the Findhelp platform we introduced to participants, could help translate goals into action, though such integrations must align with organizational workflows and participant trust.
More broadly, community-facing envisioning systems require socially situated evaluation, including participatory definitions of success and longitudinal accounts of how tools contribute to care and coordination.

%% file: implications.tex
\subsection{Design Implications for Future-Envisioning and Planning Tools}
\label{implications}

Our findings point toward design principles for future-envisioning and planning tools in contexts of constraint, instability, and institutional dependence. 
While our study focused on reentry and recovery, these implications may extend to other settings where planning requires ongoing coordination with external actors, rules, and relationships.
We argue that future envisioning should not be treated as an individual exercise or organized around a singular plan. 
Instead, future envisioning and planning should 
(1) center maintenance work, 
(2) treat revision as continued engagement, and 
(3) anticipate and support mediated use.

\subsubsection{Center Maintenance Work}

Future-envisioning tools typically foreground attainment, as seen in their focus on eliciting what users want to get or become. 
Our findings suggest that in constrained settings, what must be sustained is equally central.
Participants did not describe visions as standalone milestones.
They saw them as configurations of interdependent conditions, where sobriety, housing stability, family relationships, documentation, and program compliance mutually shaped what was feasible.
Planning, in practice, required making this maintenance work visible and actionable.

This leads to three design considerations.
First, tools can distinguish between attainment goals (acquiring something new) and sustaining goals (maintaining something that exists).
This requires eliciting and appreciating the assets that are already available to users (e.g., \cite{wong2021reflections, gautam2020crafting}). 
Second, tools can elicit dependencies and threats as a central planning scaffold. 
In our study, naming constraints enabled participants to reason concretely about their pathways and where support was necessary.
Third, progress indicators should be designed to treat stability under precarious conditions as success.
Remaining housed, sustaining employment, or maintaining sobriety is not stasis. 
It is active work that tools should help users see, track, and value.

\subsubsection{Treat Revision as Continued Engagement}
Many planning tools treat the initial plan as the artifact of value, with changes implicitly signaling deviation or failure. 
Our findings suggest the opposite.
Revision helped participants sustain engagement with possible futures as circumstances shifted.
The ability to return to earlier boards, update priorities, and hold multiple trajectories in parallel kept planning viable.

For design, this means revision should be treated as a primary interaction, not a corrective action. 
To begin with, language on the interface matters.
Terms such as ``finalize,'' ``complete,'' or ``lock in'' imply resolution, whereas  ``save version,'' ``revisit,'' or ``evolve'' signal that change is expected.
Meaningful version histories enable users to see continuity across change and to recognize learning and resilience even when aspirations shift.
Tools can also support parallel futures by allowing multiple boards or branches instead of forcing premature commitment to a single narrative.

This orientation toward evolving futures aligns with repair-centered perspectives that emphasize the value of maintenance work \cite{jackson2014repair, houston2016values}.
We advocate for designs that recognize and support the labor required to keep a future viable.

\subsubsection{Support Mediated Use}

Future-envisioning tools are often designed for individuals.
Our findings suggest that planning is collaborative and mediated.
Participants' goals depended on staff assistance, program connections, and coordination with external services, and facilitation shaped how they navigated the tool.
FutureBound's value lies in its integration into a care setting,  not in individual engagement. 
We echo HCI scholarship that advocates making institutional relationships, constraints, and eligibility requirements visible for intermediated action \cite{simonsen2020infrastructuring, karusala2017care, sambavisan2010intermediated}. 

Designers should therefore treat organizations and staff as intended users and the relationships and routines of care as fundamental to the designed tool.
Tools should be designed as components within constellations of support \cite{gatz2025node}.
Concretely, this may imply shared views that enable staff and participants to examine boards together, exportable outputs that incorporate existing documentation and handoff practices, and conversation scaffolds for joint sensemaking.
They can also include staff-oriented affordances that build organizational capacity without turning the system into surveillance \cite{mcerlean2025examining}, such as flagging dependency barriers, tracking follow-ups, or curating local resource options.

Designing for mediation also requires attention to organizational boundaries. 
Tools should support continuity of care without implying that technology can substitute for relational support or expand what organizations can provide \cite{tseng2022care}.
The design objective is not to automate planning, but to strengthen the relationships and coordination through which planning becomes actionable. 
This requires shared interpretation, coordinated next steps, and sustained support across changing conditions. \\

Staff feedback echoed these arguments. A program lead valued the tool's support for continuity and revisability, noting, ``\textit{I love that you can go back and edit it ... because you can change your vision board as your life changes, as your ideas change.}''
Staff also flagged gaps in image representation across skin tones and family structures, a reminder that tools embedded in care settings must give organizations enough control over content to avoid reproducing exclusion or narrowing which futures feel possible.

%% file: conclusion.tex
\section{Conclusion}
\label{conclusion}

We examined how community members navigating reentry used future-envisioning practices, including our digital tool, to articulate current circumstances, desired futures, and pathways forward. 
Across three workshops with 18 participants and follow-up interviews with 10 participants, we found that envisioning supported reflection on constraints and possibilities while helping participants surface assets, identify needs, and reason about dependencies within fragmented institutional systems.

Rather than treating progress as a steady march toward stable endpoints, participants described reentry and recovery as the ongoing work of keeping plans viable. 
Revising goals was a form of agency through which they maintained fragile arrangements as conditions shifted, and not a sign of failure. 
We draw design implications for future-envisioning tools and argue for systems that make maintenance visible, normalize revision, and support mediated use across the care ecologies that shape which futures become possible.

We end with a question for ourselves and for the field: do dominant ways of thinking about empowerment and progress devalue the hidden labor that goes into ``just staying okay''? 
Following \citet{jackson2014repair} and others, we suggest that designing for maintenance may be one of the most meaningful contributions HCI can make in an increasingly fragmented world.